\documentstyle[twoside,fleqn,espcrc2,epsf]{article}

\newcommand{\AmS}{{\protect\the\textfont2
  A\kern-.1667em\lower.5ex\hbox{M}\kern-.125emS}}

\title{
Lattice QCD at finite temperature 
}

\author{ Shinji Ejiri 
\address{Center for Computational Physics, 
University of Tsukuba, Tsukuba, Ibaraki 305-8577, Japan}}

\begin{document}

\begin{abstract}
Recent developments in finite-temperature QCD with dynamical quarks 
are reviewed 
focusing on the topics of critical temperature, the equation of state, 
and critical behaviors around the chiral phase transition. 
\end{abstract}

\maketitle

\section{Introduction}
\label{sec:intro}

In the last years, much effort has been paid to experimentally detect 
the quark-gluon plasma (QGP) in heavy-ion collisions. 
In these experiments, precise theoretical inputs 
about the nature of the transition 
and the equation of state (EOS) of QGP 
are indispensable to unambiguously identify the signal of QGP. 
%from the data of heavy-ion collisions in which several thousands of 
%hadrons are produced.
%
Here, lattice QCD provides us with the most powerful method 
to compute thermodynamic properties of QGP from the first principles 
of QCD.

In the quenched approximation of QCD, 
the study of finite temperature QCD is
already well matured:
Computations of $T_c$ and EOS have been made 
on lattices with several temporal lattice sizes $N_t$,
using both the standard plaquette action and various improved actions. 
Continuum extrapolation to the limit of large $N_t$ has been performed 
with different actions to confirm good agreement 
within an accuracy of a few percent\cite{Boy96,Bei99,okamoto}. 
For the pressure, precise comparisons of the results from different 
calculation methods, i.e.\, the integral method\cite{Eng90} and 
the derivative method\cite{Eng82}, 
have also been made\cite{Kla98,Eji98,Eng99}.
The expected absence of the pressure gap at the first order deconfinement 
transition point
%which is a non-trivial issue in the derivative method, 
has also been demonstrated through a new non-perturbative calculation of 
anisotropy coefficients\cite{Eji98}. 

The quenched study is now entering a new stage: 
%Temperature dependence of hadron masses and widths are important 
%in a phenomenological investigation of QGP. 
Thermal effects on the pole mass have been studied by the QCDTARO 
collaboration using an anisotropic lattice\cite{Taro00}. 
Several groups are also trying to compute hadron spectral functions 
at finite-temperature applying the maximum entropy 
method\cite{Taro97,Asa99,Wet00,Oev00}. 

On the other hand, the study of full QCD at finite temperatures 
is still at the stage of development. 
Until recently, EOS with dynamical quarks has been computed only for 
the staggered fermion.
The MILC collaboration calculated EOS in 2 flavor QCD with the 
standard staggered fermion on $N_t=4$ and 6 lattices\cite{Ber97b}.
The Bielefeld group computed EOS for $N_f=4$ using 
%a Symanzik-improved gauge action combined with 
an improved staggered fermion at $N_t=4$\cite{Eng97}. 
This year, the Bielefeld group extended the computation 
to $N_f=2$ and 3 and studied the $N_f$-dependence\cite{Kar00}. 

A new development of this year is the first calculation 
of EOS for a Wilson-type fermion\cite{CPPACS00b}. 
Using a renormalization group (RG) improved gauge action\cite{iwasaki} 
coupled with 2 flavors of clover-improved Wilson quarks\cite{SWclover}, 
the CP-PACS collaboration performed a systematic study of the pressure 
and the energy density of QGP.

In this report, we concentrate on the development in finite temperature
QCD with dynamical quarks, focusing on the topics of EOS, 
and attempt a comparison of the results from different lattice fermions.
In section \ref{sec:phase}, we discuss about the phase structure, 
critical properties around the chiral phase transition of 2 flavor QCD, 
and the transition temperature. 
we then discuss new calculations of EOS in section~\ref{sec:EOS}.
A brief summary is given in section~\ref{sec:summary}.

\section{Phase structure}
\label{sec:phase}

\begin{figure}
\vspace*{-5mm}
\centerline{
\epsfxsize=8.0cm\epsfbox{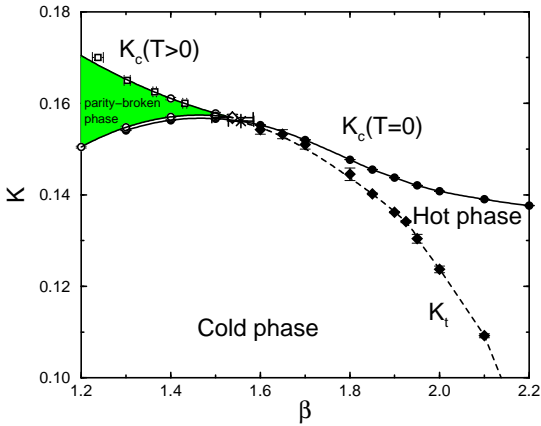}
}
\vspace*{-13mm}
\caption{
Phase diagram for the RG-improved gauge action and the clover-improved 
Wilson quark action on an $N_t=4$ lattice\cite{CPPACS00a}.
}
\vspace*{-2mm}
\label{fig:phase}
\end{figure}

\subsection{Phase structure with Wilson fermions}

It has been difficult to study EOS with Wilson-type quarks
due to several difficulties:
One is the complicated phase structure due to the explicit 
violation of chiral symmetry and due to the appearance 
of the parity-flavor broken phase (Aoki phase)\cite{aoki,saoki}. 
Therefore, a systematic study surveying a wide range of the parameter 
space is required. 
The second reason is that,
when the standard plaquette gauge action and the standard Wilson 
quark action are used, 
lattice artifacts are large on coarse lattices used in most 
finite temperature simulations, 
For example, unexpected strong phase transition is observed 
at intermediate quark masses in 2 flavor QCD\cite{Ber92,Iwa96}.
Therefore, we have to use improved actions to suppress these 
lattice artifacts. 

The CP-PACS collaboration studied 2 flavor QCD 
with an RG-improved gauge action combined with a clover-improved 
Wilson quark action\cite{CPPACS00a}. 
The first step is to clarify the phase structure for this combination 
of actions.

Figure~\ref{fig:phase} is the result for the phase diagram of QCD 
obtained on an $N_t=4$ lattice. 
The solid line $K_c(T=0)$ is the location of the chilal limit 
at zero temperature. 
The pion mass decreases as $K$ is increased from small $K$, 
and vanishes on the line $K_c(T=0)$. 
When $N_t$ is fixed as in most simulations, 
%because the lattice spacing $a$ becomes smaller as $\beta$ is increased, 
the temperature $T=(N_t a)^{-1}$ becomes higher as $\beta$ is increased. 
The dashed line $K_t$ is the pseudo-critical line for the finite temperature 
phase transition for $N_t=4$.
The region to the right of $K_t$ (larger $\beta$) is the high 
temperature QGP phase, and that to the left of $K_t$ (smaller $\beta$) is 
the low temperature hadron phase. 
The crossing point of the $K_c(T=0)$ and the $K_t$ lines is the chiral phase 
transition point\cite{Iwa96}. 

In the region above the $K_c(T=0)$ line, a parity-flavor symmetry 
of the Wilson-type quark action is broken spontaneously\cite{aoki,saoki}. 
At zero temperature, the boundary of the parity-flavor broken phase 
(the $K_c(T=0)$ line) is known to form a sharp cusp touching the free 
massless fermion point $K=1/8$ at $\beta=\infty$.

At finite temperatures, the parity-flavor broken phase retracts from the 
large $\beta$ region.
The boundary of the parity-flavor broken phase at $T>0$
(the $K_c(T>0)$ line) is shown in Fig.~\ref{fig:phase} for $N_t=4$. 
As shown in this figure, the chiral transition point locates close to 
the cusp of the parity-flavor broken phase, the difference being 
at most of $O(a)$.
This is consistent with the picture that the massless pion, i.e.\ 
the Goldstone boson associated with spontaneous chiral symmetry breaking,
appears only in the cold phase. 

\begin{figure}[t]
\vspace*{-5mm}
\centerline{
\epsfxsize=8.0cm\epsfbox{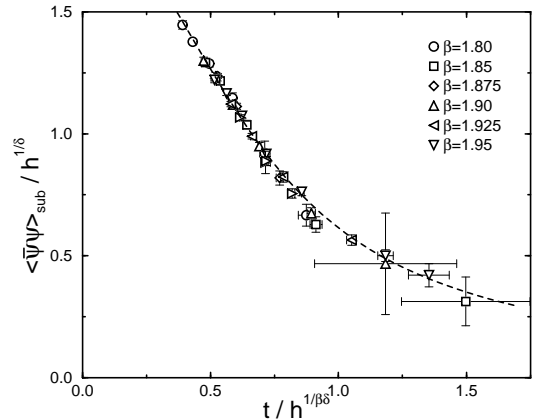}
}
\vspace*{-13mm}
\caption{
O(4) scaling relation measured on a $16 \times 4$ lattice 
with the RG gauge and clover quark actions\cite{CPPACS00a}.
}
\label{fig:o4}
\vspace*{-2mm}
\end{figure}

\subsection{Scaling study of 2 flavor QCD}

An important step towards clarification of the QCD transition 
in the real world is the study of the finite temperature 
chiral phase transition in $N_f=2$ QCD. 
When the chiral transition in $N_f=2$ QCD is second order, the transition 
is expected to be universal to that in a 3-dimensional O(4) spin model:
With the identifications
$M \sim \langle \bar{\Psi} \Psi \rangle$ for the magnetization, 
$h \sim m_{q} a$ for the external magnetic field
and $t \sim \beta - \beta_{ct}$ for the reduced temperature,
where $\beta_{ct}$ is the chiral transition point, 
we expect the same scaling behavior as the O(4) spin model. 
Therefore, confirmation of the O(4) scaling in QCD provides us 
with a test of the order of the chiral transition. 

The O(4) scaling was first tested with staggered fermion by Karsch and 
Laermann\cite{Kar94}. 
The study was extended to a wider range of the quark mass and 
lattice sizes by the JLQCD collaboration\cite{JL98} and 
by the Bielefeld group\cite{Bi98}. 
However, an agreement of the critical exponents between the O(4) spin model 
and QCD with 2 flavors of staggered fermions was not obtained for $N_t=4$.

For the case of Wilson fermion, 
Iwasaki {\it et al.}\cite{o4scale} investigated the scaling relation; 
\begin{eqnarray}
M / h^{1/\delta} = f (t / h^{1/\beta \delta}) \label{eq:scfn}
\end{eqnarray}
using the RG gauge action combined with the standard Wilson 
quark action. 
They identified the subtracted chiral condensate 
defined by an axial Ward-Takahashi identity\cite{bochicchio} 
\begin{eqnarray}
\langle \bar{\Psi} \Psi \rangle_{\rm sub}= 2m_q a 
  Z \sum_{x} \langle \pi (x) \pi(0) \rangle, \label{eq:pbpsub}
\end{eqnarray}
as the magnetization of the spin model.
Here, the quark mass $m_q$ is defined by an axial vector 
Ward-Takahashi identity\cite{bochicchio,itoh}, 
and the tree-level renormalization coefficient $Z = (2K)^2$ was adopted.
They found that the scaling relation (\ref{eq:scfn}) is well satisfied 
with the O(4) critical exponents and the O(4) scaling function. 

Figure~\ref{fig:o4} is the new result for the case of 
the clover-improved Wilson quark action coupled with the RG gauge action, 
obtained on a $16^3 \times 4$ lattice\cite{CPPACS00a}. 
The vertical axis is $M/h^{1/\delta}$ and the horizontal axis is 
$t/h^{1/\beta \delta}$, 
where $\beta$ and $\delta$ are the critical exponent obtained 
in the O(4) spin model\cite{kaya}. 
The dashed line is the O(4) scaling function\cite{toussaint}. 
They fitted the data to the scaling function adjusting $\beta_{ct}$ and 
the scales of two axes.
As seen from Fig.~\ref{fig:o4}, 
QCD data is well described by the O(4) scaling ansatz also for this 
combination of actions. 
The best fit gives $\beta_{ct}=1.469(73)$ for the chiral transition point, 
with $\chi^{2}/N_{DF}=0.82$.
This result suggests that the chiral transition is of second order 
for $N_f=2$.

This year, the MILC collaboration published their final results of 
a study of scaling properties in QCD with standard staggered 
quarks\cite{Ber00}.
They found that the QCD data are inconsistent with the O(4) scaling,
in accord with the previous results for the critical exponents\cite{Bi98,JL98}.
At the same time, a comparison of the results for $N_t=4$--12 suggests 
that the discrepancy with the O(4) scaling function 
may be removed when $N_t$ is sufficiently large. 

A different approach to clarify the origin of the unexpected results was 
proposed by Kogut, Laga\"{e} and Sinclair\cite{Kog00}.
They carried out a simulation at precisely zero quark mass 
by adding the additional, irrelevant four-fermi interaction to QCD. 
From their results obtained on $N_t=4$ and 6 lattices, 
they also suggested importance of simulations at larger $N_t$.

For staggered quarks, a comparison with an O(2) spin model is 
also important because the symmetry of the staggered quark action is 
O(2) at finite lattice spacings. 
Engels {\it et al.} calculated the scaling function for the O(2) spin 
model\cite{Eng00}.
They found that the O(2) scaling function is quite similar to the 
O(4) scaling function. 
This implies that the O(2) scaling also cannot explain the behavior of
the staggered quarks at small $N_t$.

\begin{table*}[tb]
\caption{Chiral transition temperature $T_c$ in physical units.
The scale is fixed by $\sqrt{\sigma} \simeq 425 {\rm MeV}$ for quenched 
$(N_f=0)$ QCD, and by 
$m_V = 770 {\rm MeV}$ for $N_f=2$ and $3$ QCD.
}
\label{tab:tc}
\begin{center}
\begin{tabular}{ccccccc}
\hline
action & Ref. &$N_f$ & $N_t$ & $T_c / \sqrt{\sigma}$ & $T_c / m_{V}$ 
& $T_c {\rm (MeV)}$ \\
\hline
plaquette & \cite{Bei99,Edw98}  & 0 & $\infty$ &  0.630(5) &  & 268 \\
Symanzik  & \cite{Bei99}        & 0 & $\infty$ &  0.634(4) &  & 269 \\
RG        & \cite{okamoto}      & 0 & $\infty$ &  0.650(5) &  & 276 \\
RG+clover   & \cite{CPPACS00a} & 2 & 4 &            & 0.222(5)  & 171(4) \\
Symanzik+p4 & \cite{Karsch}    & 2 & 4 &  0.425(15) & 0.225(10) & 173(8) \\
Symanzik+p4 & \cite{Karsch}    & 3 & 4 &            & 0.20(1)   & 154(8) \\
\hline
\end{tabular}
\vspace*{-4mm}
\end{center}
\end{table*}

\begin{figure}[t]
\vspace*{-5mm}
\centerline{
\epsfxsize=7.6cm\epsfbox{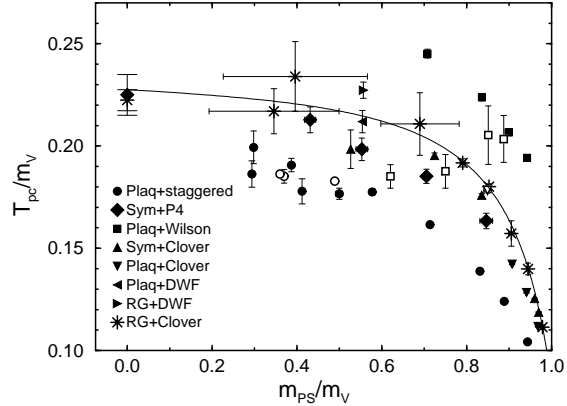}
}
\vspace*{-13mm}
\caption{
$T_{pc}/m_{V}$ as a function of $m_{PS}/m_{V}$ with various actions. 
}
\label{fig:tcmrho}
\vspace*{-2mm}
\end{figure}

\begin{figure}[t]
\vspace*{-3mm}
\centerline{
\epsfxsize=7.5cm\epsfbox{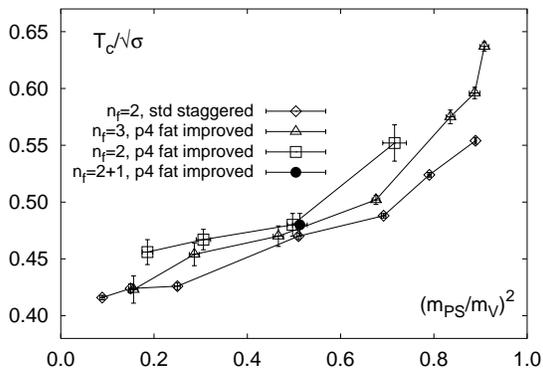}
}
\vspace*{-8mm}
\caption{
Pseudo-critical temperature $T_{pc}$ in a unit of $\sqrt{\sigma}$ for 2, 3, 2+1 flavor QCD using the p4-improved staggered action\cite{Karsch}.
}
\vspace*{-2mm}
\label{fig:tcKS}
\end{figure}

\subsection{Transition temperature}

In Fig.~\ref{fig:tcmrho}, we summarize the results of the pseudo-critical 
transition temperature $T_{pc}$ in $N_f=2$ QCD with various actions;
plaquette gauge combined with the standard staggered and Wilson 
quark\cite{karsch99,Ber97}, 
Symanzik gauge with p4-improved staggered quark\cite{Kar00,Karsch}, 
plaquette gauge with clover quark\cite{Edw99}, 
Symanik gauge with clover quark\cite{Ber97}, 
RG gauge with clover quark\cite{CPPACS00a}, 
and plaquette and RG gauge with domain-wall quark\cite{Vra99}.
The scale is fixed by the vector meson mass $m_V$ at the simulation point.
Filled and open symbols are respectively the results on $N_t=4$ and 6 lattices. 
Diamonds and stars are new; 
the diamonds from the p4-improved staggered quark by the Bielefeld group\cite{Karsch}, 
and the stars from the RG gauge with clover quark 
by the CP-PACS collaboration\cite{CPPACS00a}.
We note that the new results are slightly larger than the previous results 
from the standard staggered quarks presented by circles. 
The Bielefeld group and the CP-PACS collaboration also attempted 
chiral extrapolations. The results agree with each other within errors 
in the chiral limit, giving $T_c \sim 170$ MeV for the chiral transition
temperature.

The Bielefeld group also investigated the $N_f$-dependence of
$T_{pc}$ on a $16^3 \times 4$ lattice\cite{Karsch}.
Figure~\ref{fig:tcKS} shows their results for 2 (square) and 3 
(triangle) flavors of degenerate staggered quarks, 
normalized by the string tension.
%as a function of the mass ratio of pseudo-scalar and vector meson 
%$(m_{PS}/m_{V})^{2}$.
From this figure, the dependence on $N_f$ seems to be small for $N_f=2$--3. 
%Because the results for $N_f=2$ and 3 corresponds to the cases of 
%$m_s = \infty$ and 0, 
%we expect that the transition temperature for a more realistic case 
%with 2 light and one intermediate ($m_s \simeq T_c$) quarks 
%is close these results.

Table~\ref{tab:tc} is a summary of the transition temperature in the chiral 
limit. 
The quenched results are extrapolated to the continuum limit.
However, those for full QCD, $T_c \sim 150$--180 MeV, 
are the results for $N_t=4$, 
and confirmation at lager $N_t$ is required.

\begin{figure*}[t]
\vspace*{-5mm}
\centerline{
\epsfxsize=8.0cm\epsfbox{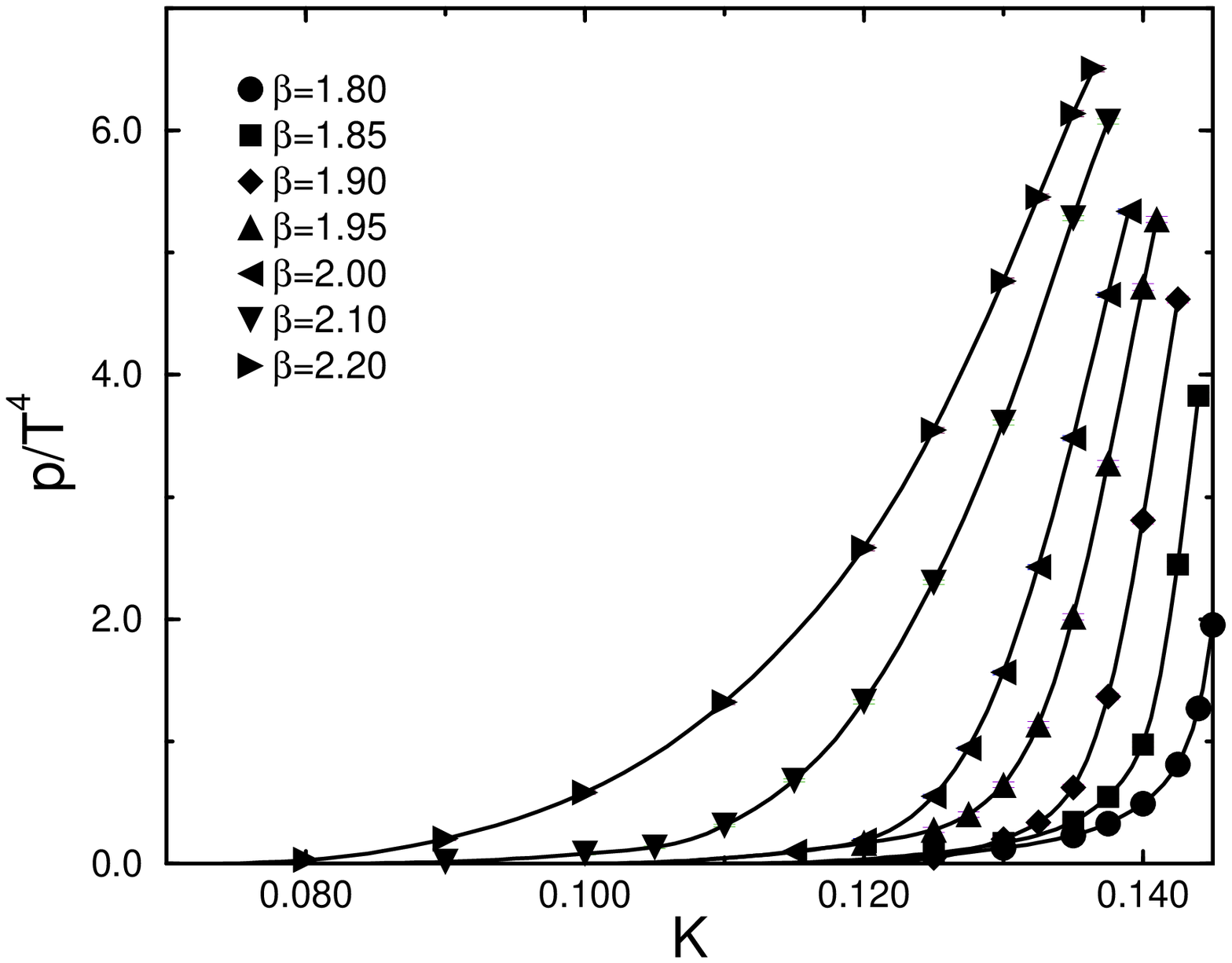}
\hspace{0mm}
\epsfxsize=8.0cm\epsfbox{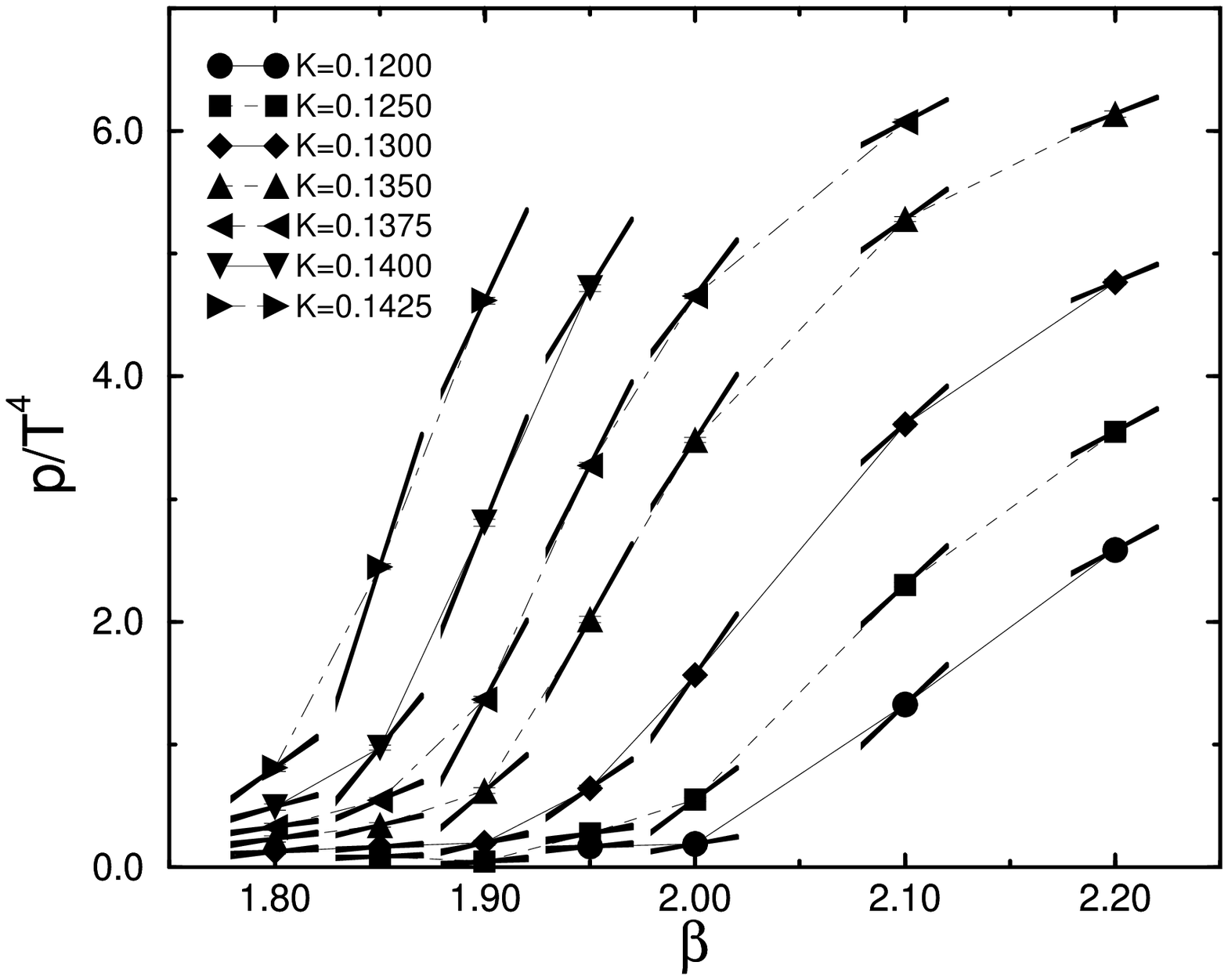}
}
\vspace*{-13mm}
\caption{
Pressure calculated with RG gauge and clover quark actions on a 
$16^3\times4$ lattice, 
as a function of $K$ (left) and $\beta$ (right).
The short lines on the symbols in the right figure express 
the slope of the pressure in the $\beta$-direction, calculated 
independently\cite{CPPACS00b}.
}
\label{fig:prs-K}
\vspace*{-2mm}
\end{figure*}

\section{Equation of state}
\label{sec:EOS}

The equation of state (EOS) is one of the most basic information 
in a phenomenological study of QGP. 
In this section, we discuss the latest results for EOS from full QCD. 
Among various methods to compute the pressure, 
the integral method\cite{Eng90} is commonly used in recent investigations. 
This method is based on the equation, $p=-f$, valid for large homogeneous 
systems,
where $f= (-T/V) {\rm ln} Z$ is the free energy density. 
Because the derivatives of the partition function can be expressed 
by expectation values of operators, 
which are computable by a Monte-Carlo simulation, 
we obtain the pressure by integrating this expectation value in the 
parameter space.
%, $(\beta, K)$ for Wilson or $(\beta, m_q a)$ for staggered. 
For the case of the Wilson quark, we have 
\begin{eqnarray} 
\frac{p}{T^4} = -\frac{f}{T^4} \!\!\! &=& \!\!\! - N_t^4 
\int^{(\beta,K)} \!\!\!\!\!\!\!{\rm d} \xi \left\{ \frac{1}{N_{s}^{3} N_{t}}
\left\langle \frac{\partial S}{\partial \xi} \right\rangle \right. 
\nonumber \\
& & \left. - ({\rm value \ at \ } T=0) 
\right\}
\end{eqnarray}
with ${\rm d} \xi=({\rm d} \beta', {\rm d} K')$ on the integration path.
The starting point of the integration path should be chosen such that 
$p \approx 0$ there.

\subsection{Equation of state with Wilson quarks}

The CP-PACS collaboration carried out a systematic calculation of EOS 
in $N_f=2$ full QCD, using the RG gauge 
and the clover quark actions on $16^3 \times 4$ and 
$16^3 \times 6$ lattices\cite{CPPACS00b}. 
Simulations at $T=0$ for subtraction and also to fix the scale and 
the line of constant physics, were made on a $16^4$ lattice. 
The derivatives of the action in $\beta$ and $K$ were computed 
using a U(1) noise method.
They found that the derivative in the $K$-direction gives a smaller error 
in the final pressure. 
Therefore, they chose the integral path in the $K$-direction. 
Figure~\ref{fig:prs-K} (left) is the $K$-dependence of $p/T^4$ for 
various $\beta$. 
In order to check the reliability of this method, they also show 
the data as a function of $\beta$ in Fig.~\ref{fig:prs-K} (right). 
The short lines on the symbols are the slope of the pressure in terms of 
$\beta$, which are measured directly from 
$\left\langle \partial S / \partial \beta \right\rangle$. 
We see that the slopes well describe the pressure curves 
which are computed from an independent measurement of 
$\left\langle \partial S/ \partial K \right\rangle$. 
This means that the value of the pressure is independent 
of the choice of the integration path in the parameter space.

\begin{figure}[t]
\vspace*{-5mm}
\centerline{
\epsfxsize=7.8cm\epsfbox{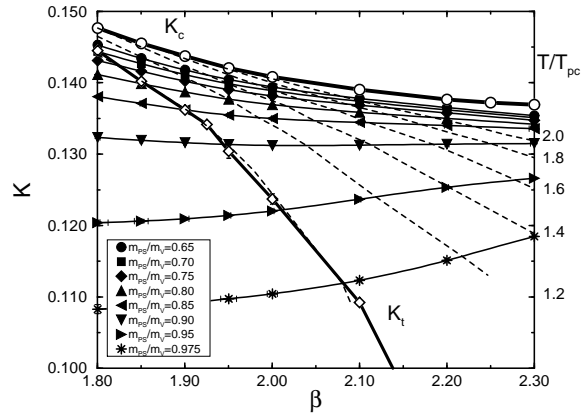}
}
\vspace*{-13mm}
\caption{
$m_{PS} / m_{V}$ constant lines and $T/T_{pc}$ constant lines for 
an $N_t=4$ lattice in the $(\beta, K)$ parameter space.
The values of $T/T_{pc}$ for the dashed lines are given on the right 
edge of the figure\cite{CPPACS00b}. 
}
\label{fig:pirhoconst}
\vspace*{-4mm}
\end{figure}

The data in this form is not yet useful for phenomenological 
applications. We wish to calculate the temperature dependence for 
each physical system, i.e., on a line of constant physics. 
The CP-PACS collaboration has chosen the condition 
$m_{PS} / m_{V}=$ constant, determined at $T=0$, to define 
the lines of constant physics. 
The relation between the simulation parameters and the physical 
parameters is then given by Fig.~\ref{fig:pirhoconst}. 
The solid lines are the lines of constant physics. 
The dashed lines are $T / T_{pc}=$ constant lines, 
where $T_{pc}$ is the pseudo-critical temperature on the same 
line of constant physics and the scale was fixed by $m_V$.
The dashed line $T/T_{pc}=1$ itself was determined by fitting the 
pseudo-critical transition points (diamonds) 
obtained on an $N_t=4$ lattice. 

\begin{figure}[t]
\vspace*{-5mm}
\centerline{
\epsfxsize=8.0cm\epsfbox{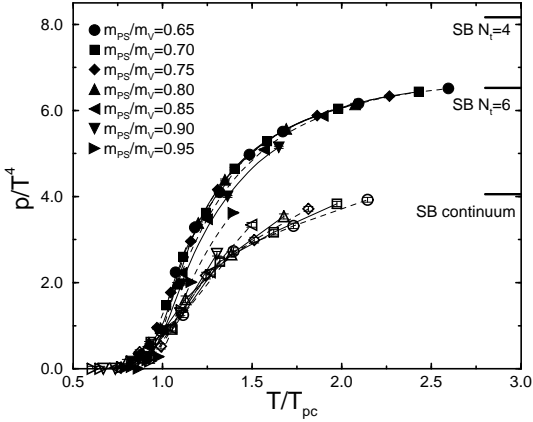}
}
\vspace*{-13mm}
\caption{
Pressure calculated with the RG gauge and clover quark actions 
on $16^3 \times 4$ and 
$16^3 \times 6$ lattices, as a function of $T/T_{pc}$\cite{CPPACS00b}.
}
\vspace*{-2mm}
\label{fig:prs-ttc}
\end{figure}

Interpolating the data in Fig.~\ref{fig:prs-K} 
yields Fig.~\ref{fig:prs-ttc} in which the pressure is given 
as a function of temperature, for each line of constant physics. 
Filled (open) symbols are the results of $N_t=4$ (6). 
Different shapes of the symbol correspond to different values of 
$m_{PS} / m_{V}$, i.e., different quark masses. 
This figure shows that the pressure is almost independent of the quark mass 
in a wide range of $m_{PS}/m_{V}$.
On the other hand, the $N_t$-dependence is sizeable: 
there exist large lattice artifacts in the $N_t=4$ results.

\begin{figure}[t]
\vspace*{-5mm}
\centerline{
\epsfxsize=8.0cm\epsfbox{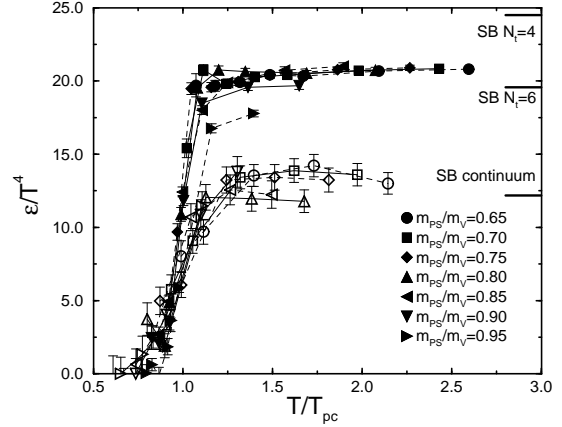}
}
\vspace*{-13mm}
\caption{
Energy density as a function of $T/T_{pc}$, 
calculated with RG gauge and clover quark actions 
on $16^3 \times 4$ and $16^3 \times 6$ lattices\cite{CPPACS00b}.
}
\vspace*{-2mm}
\label{fig:eng-ttc}
\end{figure}

The energy density is obtained by combining the results for the pressure 
and the results for the quantity;
\begin{eqnarray}
\frac{\epsilon - 3p}{T^4} \!\!\! &=& \!\!\! N_t^4\left[ \frac{1}{N_s^3 N_t} \left( 
a \frac{\partial \beta}{\partial a} 
\langle \frac{\partial S}{\partial \beta} \rangle + 
a \frac{\partial K}{\partial a} 
\langle \frac{\partial S}{\partial K} \rangle \right) \right. \nonumber \\
& & \left. - ({\rm value \ at} \ T=0) \right] . \label{e3p}
\end{eqnarray}
Here, the beta functions, $a \frac{\partial \beta}{\partial a}$ and 
$a \frac{\partial K}{\partial a}$, can be determined by measuring the 
differentials of $\beta$ and $K$ in terms of the $m_{V} a$, on each 
$m_{PS} / m_{V} =$  constant line.

Figure~\ref{fig:eng-ttc} is the final results for the energy density 
as a function of the temperature.
They indicate that the quark mass dependence for 
the energy density is also small.

Previous results from the staggered quark showed a peak just above $T_c$
on an $N_t=4$ lattice\cite{Ber97b}.
We do not see such a peak in Fig.~\ref{fig:eng-ttc} from the 
improved Wilson quark. 
Also the peak disappears at $N_t=6$ for the staggered fermion, thus, 
the peak seems to be a lattice artifact.

We find that 
the magnitude and $T$-dependence of EOS on $N_t=6$ lattices are quite 
similar between improved Wilson and staggered quarks. 
The quark mass dependence was also investigated with the staggered quark;
similar to the case of the improved Wilson quark, 
the difference between $m_q/T=0.075$ and $0.15$ is found to be 
small within errors\cite{Ber97b}.

Short lines on the right axes of Figs.~\ref{fig:prs-ttc} and 
\ref{fig:eng-ttc} are the Stefan-Boltzmann (SB) values 
%calculated from corresponding operator expectation values, 
at $N_t=4$, 6 and in the continuum limit.
Both $N_t=4$ and 6 results do not indicate a clear approach towards 
these values in the high temperature limit. 
A clear deviation from the SB limits at finite $N_t$ was also observed 
in a quenched study of EOS with the RG-improved action\cite{okamoto}. 
%using both integral and operator methods
%As a possible explanation, existence of non-perturbative contributions to 
%finite lattice corrections, due to the IR divergence of the perturbation 
%theory, was suggested. 
Further study is necessary to clarify this phenomenon.
However, quenched studies show that EOS extrapolated to the 
continuum limit $N_t=\infty$ is consistent with the continuum SB value
in the high temperature limit. 
From Figs.~\ref{fig:prs-ttc} and \ref{fig:eng-ttc}, we note that the 
results for $N_t=6$ are already close to the continuum SB limit.
An accurate continuum extrapolation may be possible already 
with data at $N_t \sim 8$ or 10.

\begin{figure}[t]
\vspace*{-4mm}
\centerline{
\epsfxsize=7.4cm\epsfbox{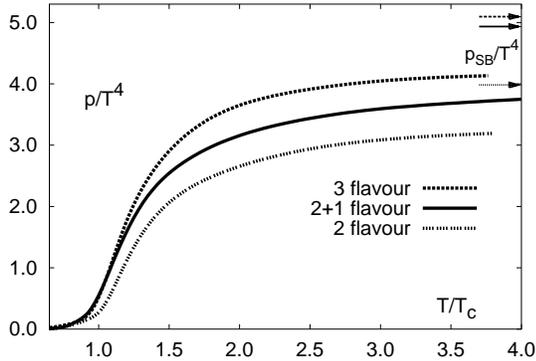}
}
\vspace*{-8mm}
\caption{
Pressure from the p4-improved staggered action on a $16^3 \times 4$ lattice 
as a function of $T/T_{pc}$\cite{Kar00}.
}
\vspace*{-4mm}
\label{fig:prsbie}
\end{figure}

\begin{figure}[t]
\vspace*{-4mm}
\centerline{
\epsfxsize=7.4cm\epsfbox{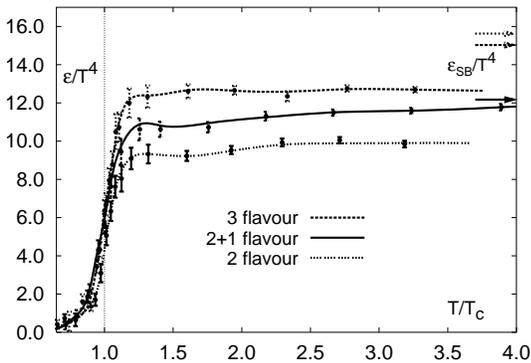}
}
\vspace*{-8mm}
\caption{
Energy density from the p4-improved staggered action on a $16^3 \times 4$ lattice 
as a function of $T/T_{pc}$\cite{Kar00}.
}
\vspace*{-4mm}
\label{fig:engbie}
\end{figure}

\subsection{$N_f$-dependence of equation of state}

So far, we have discussed simulations with 2 dynamical quarks, 
which are identified as light up and down quarks.
In this subsection, We review results on the effect of the strange quark. 
The Bielefeld group compared EOS for $N_f=2,3$ and $2+1$ on a 
$16^3 \times 4$ lattice,
using the p4-improved staggered quark action with the Symanzik improved 
gauge action\cite{Kar00}. 
$N_t$-dependence is known to be small for this action in the 
Stefan-Boltzmann limit.
They computed EOS at $m_q/T=0.4$ for $N_f=2$ and 3,  
and at $m_{u,d}/T=0.4$ and $m_s/T=1.0$ for $N_f=2+1$.

Figure~\ref{fig:prsbie} shows the results of pressure. 
A clear $N_f$-dependence is visible; 
the pressure becomes larger as $N_f$ is increased.

Clear $N_f$-dependence was observed also in the energy density. 
They also observed peaks of the energy density near $T_{pc}$, 
similar to the result of the MILC collaboration at $N_t=4$. 

They attempted to estimate the energy density in the chiral  
limit by subtracting the contribution of 
$ \langle \partial S / \partial (m_q a) \rangle $
in the calculation of the $(\epsilon -3p)/T^4$, 
since this term vanishes when the quark mass is zero. 
Quark mass dependence was assumed to be small in the pressure.
Figure~\ref{fig:engbie} shows the resulting estimate of $\epsilon/T^4$ 
in the chiral limit.
The peak disappeared in this figure. 
Therefore, in contrast to the case of Fig.~\ref{fig:eng-ttc}, 
noticeable quark mass dependence was suggested for the energy density. 
Further study is necessary to clarify this point.

\section{Summary}
\label{sec:summary}

We reviewed progress of the numerical study 
of finite-temperature QCD with dynamical quarks in the past year,
focusing on the topics of EOS.
%
%Scaling properties around the chiral transition of $N_f=2$ QCD 
%were examined. 
%In a previous study, good consistency with the expected O(4) scaling
%was reported for the standard Wilson quark combined with 
%an RG-improved gauge action. 
%This year, a further confirmation of the O(4) scaling was reported
%for a clover-improved Wilson quark. 
%
%With the standard staggered fermion, however, 
%the O(4) scaling was shown to be absent at $N_t < 12$, 
%while a better agreement was suggested at larger $N_t$.

For the transition temperature, 
recent results for $N_f=2$ QCD, 
computed with improved staggered and 
improved Wilson fermion actions on $N_t=4$ lattices, 
agree well with each other. 
In the chiral limit, the transition temperature is about 170 MeV. 
This value is higher than the previous results 
from the standard staggered fermion at $N_t=4$ and 6. 
%
%The $N_f$-dependence of the transition temperature was also studied
%with an improved staggered fermion.
%The effect of $N_f$ was found to be small between $N_f=2$ and 3. 
%
Confirmation of this result requires further study at lager $N_t$. 

The first calculation of EOS with Wilson-type fermion 
was performed for $N_f=2$ at $N_t=4$ and 6, using an improved Wilson 
fermion.
%In order to extract a prediction for the continuum limit, however, 
%further study at larger $N_t$ is required. 
The results of EOS for $N_t=6$ are quite similar between 
the improved Wilson and the staggered fermions, 
and are close to the continuum Stephan-Boltzmann value at high 
temperatures.
This may suggest that a small $N_t \sim 8$ or 10 is already 
sufficient for a precise calculation of EOS. 
On the other hand, 
%in contrast to the case of the transition temperature, 
a clear $N_f$-dependence was reported for EOS. 
This implies the importance of including dynamical strange quark 
in the simulations. 

In summary, we have obtained two keywords for the next step of 
full QCD thermodynamics: 
larger $N_t$ with improved actions towards the continuum limit, 
and simulations with the dynamical strange quark. 
Studies in these directions are now starting.

\vspace{3mm}
%\paragraph*{Acknowledgments}
I would like to thank A. Ukawa and K. Kanaya 
for helpful suggestion and comments on the manuscript. 
I also thank F. Karsch for communication on their results 
and useful discussions.
This work is supported by JSPS.

\end{document}